\documentclass[aps,pof,twocolumn,superscriptaddress,showpacs,floatfix,citesort]{revtex4}
\usepackage{epsfig}
\usepackage{amssymb}
\usepackage{amsmath}
\usepackage[dvips]{color}
\usepackage{lscape}
\usepackage{longtable}
\usepackage{rotating}
\usepackage{float}
\usepackage{bm}
\newcommand{\lp}{\left(}
\newcommand{\rp}{\right)}
\newcommand{\be}{\begin{equation}}
\newcommand{\ee}{\end{equation}}
\newcommand{\bea}{\begin{eqnarray}}
\newcommand{\eea}{\end{eqnarray}}

\def\bu{\bm{u}}
\def\bv{\bm{v}}
\def\bg{\bm{g}}
\def\bomega{\bm{\omega}}

\begin{document}
\title{Quantifying microbubble clustering in turbulent flow from single-point measurements} 

\author{Enrico Calzavarini\footnote{Present electronic address: enrico.calzavarini@ens-lyon.fr}}\address{Department of Applied Physics, 
JMBC Burgers Center for Fluid Dynamics, 
and IMPACT Institute, 
University of Twente, 7500 AE Enschede, The Netherlands.} 
  \author{Thomas H. van den Berg} \address{Department of Applied Physics, 
JMBC Burgers Center for Fluid Dynamics, 
and IMPACT Institute, 
University of Twente, 7500 AE Enschede, The Netherlands.}  
\author{Federico Toschi}
\address{IAC-CNR, Istituto per le Applicazioni del Calcolo, Viale del
  Policlinico 137, I-00161 Roma, Italy and INFN, via Saragat 1,
  I-44100 Ferrara, Italy.} 
 \author{Detlef Lohse\footnote{Electronic address: d.lohse@utwente.nl}}
 \address{Department of Applied Physics, 
JMBC Burgers Center for Fluid Dynamics, 
and IMPACT Institute, 
University of Twente, 7500 AE Enschede, The Netherlands.}

\pacs{47.55.Kf, 47.27.T-}





\date{\today}

\setcounter{page}{1}

\begin{abstract}

  Single-point hot-wire measurements in the bulk of a turbulent
channel have been performed in order to detect and quantify the
 phenomenon of preferential bubble accumulation. We show that
statistical analysis of the bubble-probe colliding-times series can
give a robust method for investigation of clustering in the bulk
regions of a turbulent flow where, due to the opacity of the flow,
no imaging technique can be employed. We demonstrate that
microbubbles ($R_0 \simeq100\ \mu m$) in a developed turbulent
flow, where the Kolmogorov length-scale is $\eta \simeq R_0$,
display preferential concentration in small scale structures with a
typical statistical signature ranging from the dissipative range,
${\cal O}(\eta)$, up to the low inertial range
${\cal O}(100\eta)$.  A comparison with Eulerian-Lagrangian numerical
simulations is also presented to further support our proposed way
to characterize clustering from temporal time series at a fixed position.
\end{abstract}

\maketitle 

\section{Introduction}
The phenomenon of preferential concentration of small particles and
bubbles in turbulent flows attracted much attention in recent years,
from experimental works \cite{bew06,cro96,fess94,douady91,bou06,hoy05,ayy06}, to numerical investigations
\cite{maxey87,elg92,elg93,maz03a,maz03b,bif05,chu05,squi91,wang93,bec06}
\footnote{
We have recently studied preferential concentration in
E. Calzavarini, M. Kerscher,  D. Lohse and  F. Toschi, {\em {Dimensionality
 and morphology of particle and bubble clusters in turbulent flow}},
 see Arxiv:0710.1705 .}, and theoretical developments
\cite{falk04}.  The preferential accumulation is
an inertial effect. 
Particles heavier than the fluid are on average ejected from
vortices, while light buoyant particles tend to accumulate in high
vorticity regions.
Small air bubbles in water (below 1 mm,
typical
Reynolds number of order ${\cal O}(1)$) can be regarded as a
particular kind of non-deformable light spherical particles with
density negligibly small compared to the fluid one. In fact, in this
size-range, shape oscillations or deformations and wake induced effect
can be reasonably neglected.  Strong preferential bubble
accumulation in core
vortex regions is therefore expected according to the inertia
mechanism, and indeed observed experimentally 
\cite{douady91} and numerically \cite{wan93b,ber06b}.
Next to the added mass force and gravity also
drag and lift forces can 
affect the clustering.
Moreover,
the coupling of the disperse phase to the fluid
flow (\textit{two-way} coupling) and the finite-size effect of
particle-particle interaction (\textit{4-way} coupling) may also
result in non-negligible factors of perturbation for preferential
concentration of particles and bubbles in highly turbulent flows.\\ \indent
Both the lift force, the two-way coupling, and the four-way coupling
are notoriously difficult to model in numerical simulations and a 
validation of the models against numerical simulations is crucial. 
However, 
experimental measurements on bubbly laden turbulent flows are
challenging, as even at very low void fractions ($\sim 1 \%$ in volume) the fluid
is completely opaque and difficult to
access with external optical methods, especially in the bulk region of
the flow.  
To experimentally explore the bubble clustering in the bulk at high
void fraction one therefore has to fall back on
\textit{intrusive}
hot-wire anemometer measurements. Such measurements had 
earlier been employed to determine  the modification
of turbulent spectra through bubbles \cite{lan91,rensen05,ramon06b}.  
For the calculation of the velocity spectra
bubbles hitting the probe had first to be identified in the hot-wire
signals 
\cite{lut05b,ren05b} and then
filtered out. In the present paper we employ the very same 
hot-wire time series to obtain information on the bubble clustering
in the turbulent flow.
An alternative method to obtain local information on the bubble
distribution may be phase doppler particle analyzers \cite{bergenblock2007}.\\
\indent
One could object that measurement from one fixed point in space are
too intrusive because they can destroy the clusters, or that they are
ineffective in extracting features of the bubble trapping in turbulent
vortical structures.  
The aim of this paper is to demonstrate that this is not the case,
when using 
appropriate statistical indicators for the analysis 
of series of
bubble colliding times on the hot-wire probe. 
We show
that it is possible to detect and quantify the micro-bubble clustering
from a one-point measurement set-up.  We compare experimental findings
with results from numerical simulations based on Eulerian-Lagrangian
approach. Due to limitations that we will discuss later, only
a qualitative agreement among numerics and experiments is
expected. Nevertheless, we show how this comparison is helpful in
clarifying the trend in the clustering at changing the turbulent
conditions.

\section{Details of the experiment methods}
The experimental set-up is the Twente water
channel, a vertical duct of square cross section with dimension $200
cm \times 45 cm \times 45 cm$. We refer to Rensen \textit{et al.}
\cite{rensen05} for a detailed description. An array of porous ceramic
plates, positioned on the top of the channel, is used to generate
co-flowing small bubbles of average radius, $R_0 \simeq 100 \mu m$, as
described in \cite{ramon06b}. Fluid turbulence is generated by means
of an active grid, positioned immediately downstream the bubble
injection sites. The typical flow is characterized by a large mean
flow, $U$, with turbulent fluctuations, $u' \equiv \langle (u_{z}(t) - U )^2 \rangle_{t}^{1/2}$, of
smaller amplitude. The condition $u'/U \ll 1$ assures that Taylor's
frozen-flow hypothesis can be applied. The dissipative Kolmogorov
scale measures typically $\eta = 400\div 500\ \mu m$, while the Taylor
micro-scale and the integral one, are respectively $\lambda \simeq 30\
\eta$, and $L_0 \simeq 500\ \eta$.
The typical bubble size is of the same order, or slightly smaller, than $\eta$.\\ \indent
We consider microbubble signals extracted from a hot-film anemometry
probe (300 $\mu m$ in size) fixed at the center of the
channel. Detection of micro-bubbles is less ambiguous than for
large bubbles where probe piercing and break-up events are highly
probable \cite{zenit01}. A micro-bubble hitting the probe produces a
signal with a clear spike. The bubble can be identified by
thresholding of the velocity time-derivative signal, see
Fig. 2 of ref. \cite{ramon06b}.
This identification procedure leads to the definition of a minimal
cut-off time in the capability to detect clustered events, two
consecutive bubbles in our records cannot have a separation time
smaller than $\tau = 10^{-2} sec$. Such dead-time is mainly linked to
the typical response-time of the acquisition set-up.  Here we consider
two time series of microbubble measurements, i.e.  hitting times,
selected from a larger database because of their uniformity and
relevant statistics. We will refer to them in the following as sample
(a) and (b). The first sample (a) has been taken for a 12 hours long
measurement; it consists of $N_b = 24099 $ bubbles with a mean hitting
frequency $f = 0.56\ sec^{-1}$. The second sample, (b), is a record of
11 hours, $N_b= 11194$ and $f \simeq 0.28\ sec^{-1}$. There are two
main differences among the experimental conditions in which the two
samples have been recorded, that is the total volume air fraction (called void fraction $\alpha$), 
and the amplitude of the mean flow and therefore the
intensity of turbulence. Case (a) has a void fraction of $\approx
0.3\%$ and (b) has instead $\alpha \approx 0.1\%$.  
Note that, even at these very small void fractions, the mean number density of bubbles
amounts to $O(10^2)$ per cubic centimeter. This explains the optical opacity of the bulk region of our system. 
Nevertheless, given the small effect produced by the dispersed bubbly phase on the turbulent
cascading mechanism \cite{ramon06b}, we consider the discrepancy in
$\alpha$ as irrelevant for the velocity spectra. In contrast, the
difference in the forcing amplitude is more important, because it sensibly
changes all the relevant scales of turbulence as summarized
in Table \ref{tab1}.
\begin{table*}
\begin{center}
\begin{tabular}{|c|c|c|c|c|c|c|c|c||c|c|c|c|}
  \hline
  & $L_0$ $(cm)$ & $U$ $(cm/s)$ & $u'$  $(cm/s)$&$Re_{\lambda}$ &$\tau_{eddy}$ $(s)$ & $\tau_{\eta}$ $(ms)$
  & $\eta$ $(\mu m)$& $u_{\eta} $ $(mm/s)$ &$Re_b$ & $R_0/\eta$ & $St$  & $g\tau_b/u_{\eta}$  \\
  \hline
  a)& 22.6 &19.4  & 1.88 & 206 & 12.0 &$151.$ & 388. & $2.57$ & 4.4 & 0.26 & 0.007  & 4.2 \\
  \hline
  (b)& 23.1 &14.2 & 1.39 & 180 & 16.6 &$240.$ & 489. & $2.04$   & 4.4 & 0.20 & 0.004   & 5.3 \\
  \hline  
\end{tabular}
\end{center}
\caption{Relevant turbulent scales and bubble characteristics for the
  two experimental samples analyzed. Fluid turbulent quantities have
  been estimated from one-dimensional energy spectra. 
  From left to right: integral scale, $L_0$, mean velocity, $U$, single-component root mean square velocity, $u'$,
  Taylor Reynolds number, $Re_{\lambda}$, large eddy turnover time, $\tau_{eddy}$, dissipative time ($\tau_{\eta}$) space ($\eta$) and velocity ($u_{\eta}$) scales, bubble Reynolds number, $Re_b$, bubble-radius and Kolmogorov-length ratio, $R_0/\eta$, Stokes number, $St$, ratio between terminal velocity in still fluid and dissipative velocity scale, $g \tau_b/u_{\eta}$.
  }\label{tab1}
\end{table*}
In particular, this leads to different values for the minimal
experimentally detectable scale: $\Delta r_{min} \simeq 5 \eta $ for
case (a) and $\Delta r_{min} \simeq 3 \eta $ for (b), where Taylor
hypothesis has been used to convert time to space measurements,
i.e. $\Delta r = \tau\cdot U$.  In the following, results of our
analysis will be presented by adopting space units made dimensionless
by the Kolmogorov scale $\eta$. We consider this re-scaling more
useful for comparison with different experiments and simulations where
a mean flow may be absent.

\section{Description of the statistical tools}
In this section we introduce the statistical tests that we will adopt
to quantify the clustering.  Due to the fact that the experimental
recording is a temporal series of events, we have necessarily to focus
on a tool capable to identify, from within this one dimensional serie,
possible signatures of 3D inhomogeneities.\\ \indent
A first way to assess the presence of preferential concentrations in
the experimental records is to compute the probability density
function (pdf) of the distance, $\Delta r$, between two consecutive
bubbles.  Whether the particles distribute homogeneously in space, their
distribution would be a Poissonian distribution and hence the distance
between two consecutive bubbles would be given by the well know
exponential expression: $\rho \exp(-\rho \Delta r)$, where $\rho = f/
U$ is the number of bubbles per unit length (i.e. their density)
\cite{feller}.
Due to the presence of turbulence we expect that in general the
spatial distribution of the bubbles will differ from a
Poissonian distribution: in any case it is natural to expect that for
separation-scales large enough the exponential form of the pdf should
be recovered.  In fact, pairs of successive bubbles with large
separations $\Delta r$, larger then any structures in the flow, are
expected to be uncorrelated, memory-less, events.\\ \indent
Due to the possible accumulation on small scales (clustering of
bubbles) the long tail of the pdf may have an exponential decay rate
that is different from the global mean, $\rho$.  The tail of the
experimentally measured pdf can be fitted with an exponentially
decaying function, $A\cdot \exp(-\rho_h \Delta r)$, with a rate that
we call $\rho_h$, where $h$ stands for homogeneous. In case of
small-scale clustering we expect $\rho_h$ to be smaller than $\rho$.
As an indicator of the fraction of bubbles accumulated in turbulent
structures, we use the coefficient $\mathcal{C} \equiv 1 -
\rho_h/\rho$, whose value varies in between 0 and 1.\\ \indent
The test so far introduced is useful but only provides an indication
on how homogeneously distributed the bubbles are 
at small scales, while it gives no indication on their possible
``large-scale'' correlations.  Here we introduce a second, more
comprehensive, statistical test particularly convenient to reveal the
scales at which the inhomogeneity develops.  The idea is
to compute the coarse-grained central moments of the number of
bubbles, on a window of variable length $r$, $\mu^{p}_{r} \equiv
\langle ( n - \langle n \rangle_{r} )^p \rangle_{r}$. The length of
the window $r$ will be the scale at which we study whether the
distribution resembles  a homogeneous one.  We will focus
on scale dependent kurtosis and skewness excesses, respectively: $K(r)
\equiv \mu^{ 4}_{r}/(\mu^{ 2}_{r})^2 - 3$ and $S(r) \equiv \mu^{
  3}_{r}/(\mu^{ 2}_{r})^{3/2}$.  A random distribution of particles
spatially homogeneous with mean density $\rho$  corresponds to
the Poissonian distribution : $p(n) = \exp (- \rho\ r) (\rho r)^n
(n!)^{-1}$, where $r$ is the length of the spatial window and $n$ is
the number of expected events.  Therefore, once the particle space
rate $\rho$ is given, the value of any statistical moment
can be derived for the corresponding window length $r$. A
spatially Poissonian distribution of particles implies the functional
dependences $K(r) = (\rho r)^{-1}$ and $S(r) = (\rho r)^{-1/2}$ .
Furthermore, we note that at the smallest scale, when $r= \Delta
r_{min}$, we reach the singular limit (shot-noise limit) where for any
given space-window we can find none or only one bubble and all
statistical moments collapse to the same value. This latter limit,
which is by the way coincident with Poisson statistics, represents our
minimal detectable scale.  We are interested in departures from the
shot-noise/random-homogeneous behavior for the statistical observables
$K(r)$ and $S(r)$.
\begin{figure}
\begin{center}
\vspace*{0.4cm}
\epsfig{file=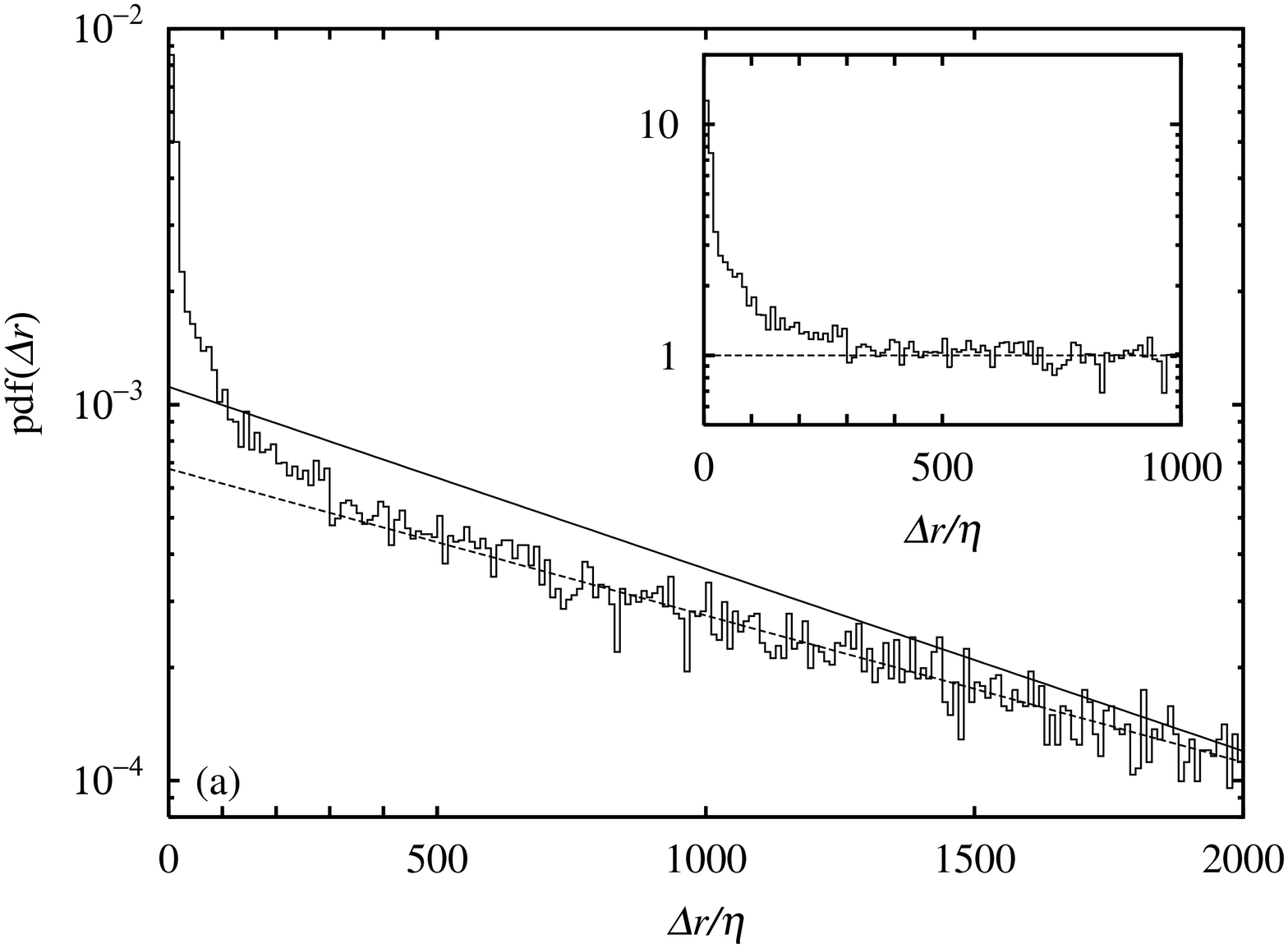,width=.3\textwidth,angle=-90}\\
\vspace*{0.6cm}
\epsfig{file=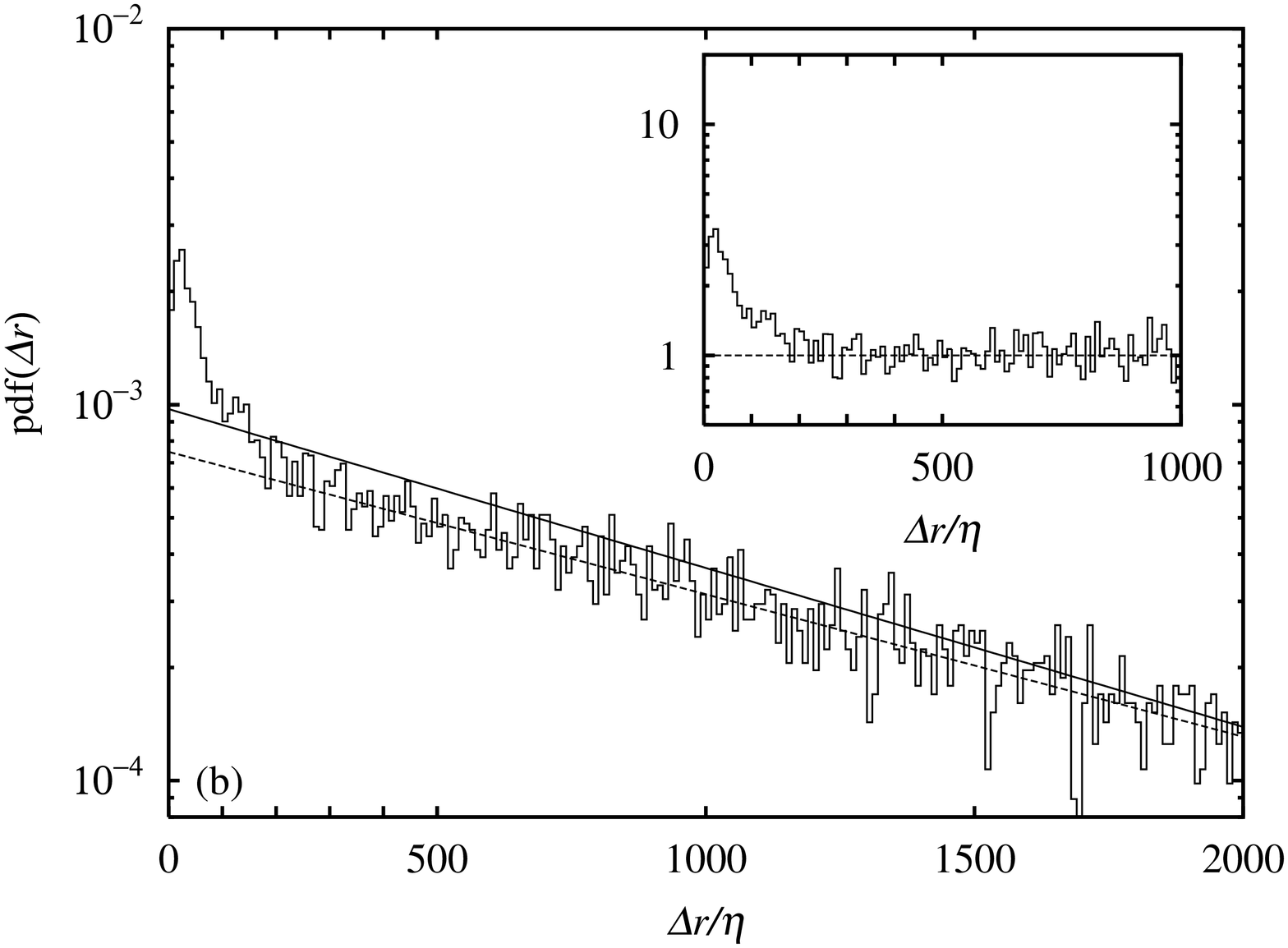,width=.3\textwidth,angle=-90}
\vspace*{0.3cm}
\end{center}
\caption{
Probability density function of distance between successive bubbles, pdf($\Delta r$). Exponential
behavior, $\rho e^{-\rho \Delta r}$, (solid line) and exponential
fit, $A \cdot e^{-\rho_h \Delta r}$, of the large-scale tail (dashed
line) are reported. The inset shows the pdf($\Delta r$) compensated
by the fitted large-scale exponential behavior, i.e. the pdf($\Delta r$) divided by $A \cdot e^{-\rho_h \Delta r}$.}\label{fig1}
\end{figure}
\section{Results of the analysis on experimental data}

In figure \ref{fig1} we show the computed pdf$(\Delta r)$ for the two
data samples considered. Deviations from global homogeneity are clear
if the shape of the histogram is compared with the solid line
representing the pdf $\rho \exp(-\rho \Delta r)$. These deviations are
slightly more pronounced in the more turbulent case (a)
as compared to 
case (b).
Nevertheless, one can notice that the pure exponentially decaying
behavior, i.e. homogeneity, is recovered from distances of the order
of ${\cal O}(100 \eta)$ up to the large scales. The dotted line on
Fig. \ref{fig1}, which represents the linear fit on the long
homogeneous tail in the interval $[10^3, 2\cdot 10^3] \eta$, and the
inset boxes, where the pdf is compensated by the fit, shows this
latter feature.  The evaluation of the coefficient $\mathcal{C}$ leads
to values for the relative bubbles excess in clusters corresponding to
19 \% for case (a) ($Re_{\lambda} \simeq 206$) and 10 \% for case (b)
($Re_{\lambda} \simeq 180$), confirming the trend of stronger
concentration in flows with stronger turbulence level.
\begin{figure}
\begin{center}
\vspace*{0.4cm}
\epsfig{file=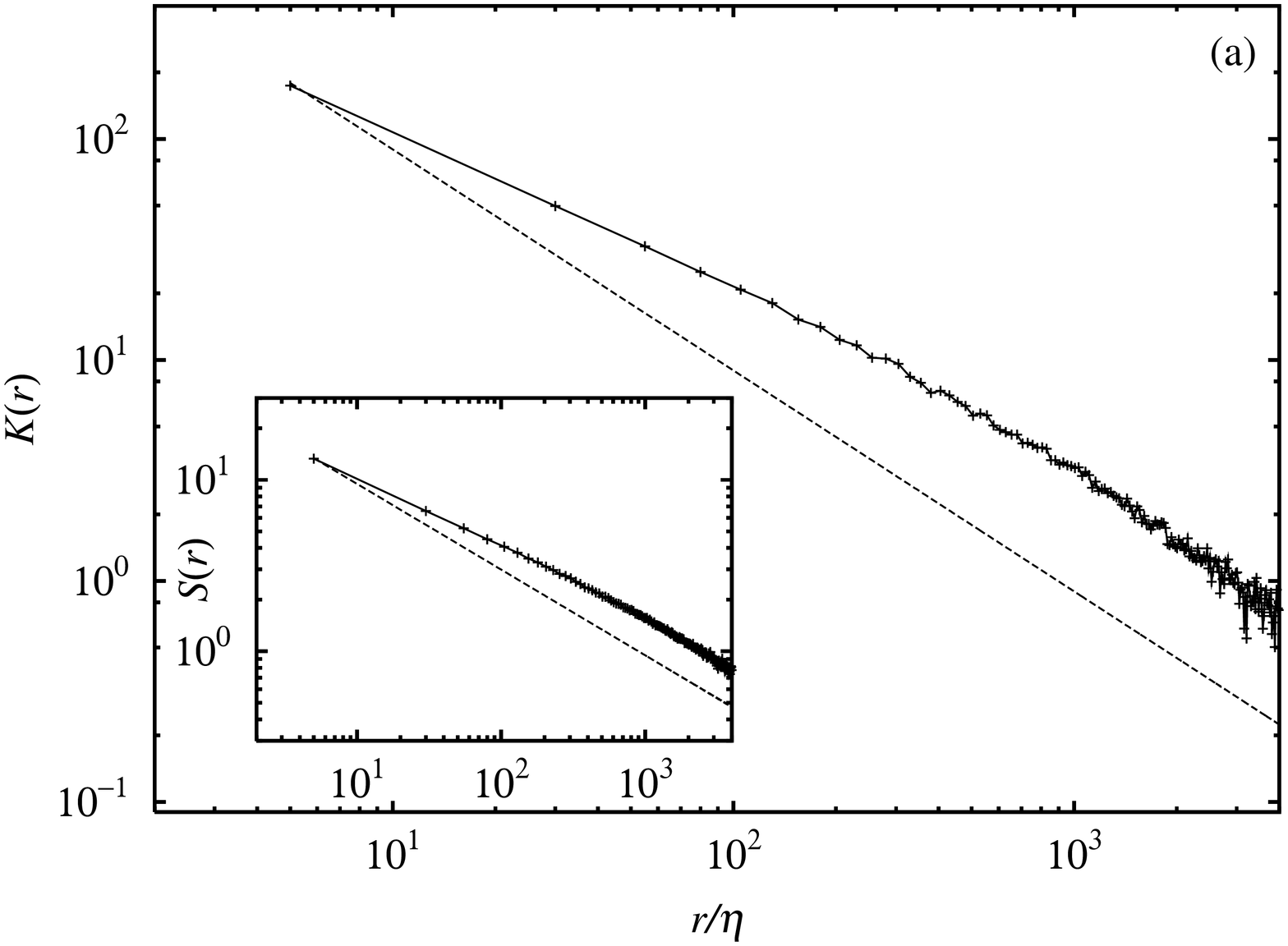,width=.3\textwidth,angle=-90}\\
\vspace*{0.6cm}
\epsfig{file=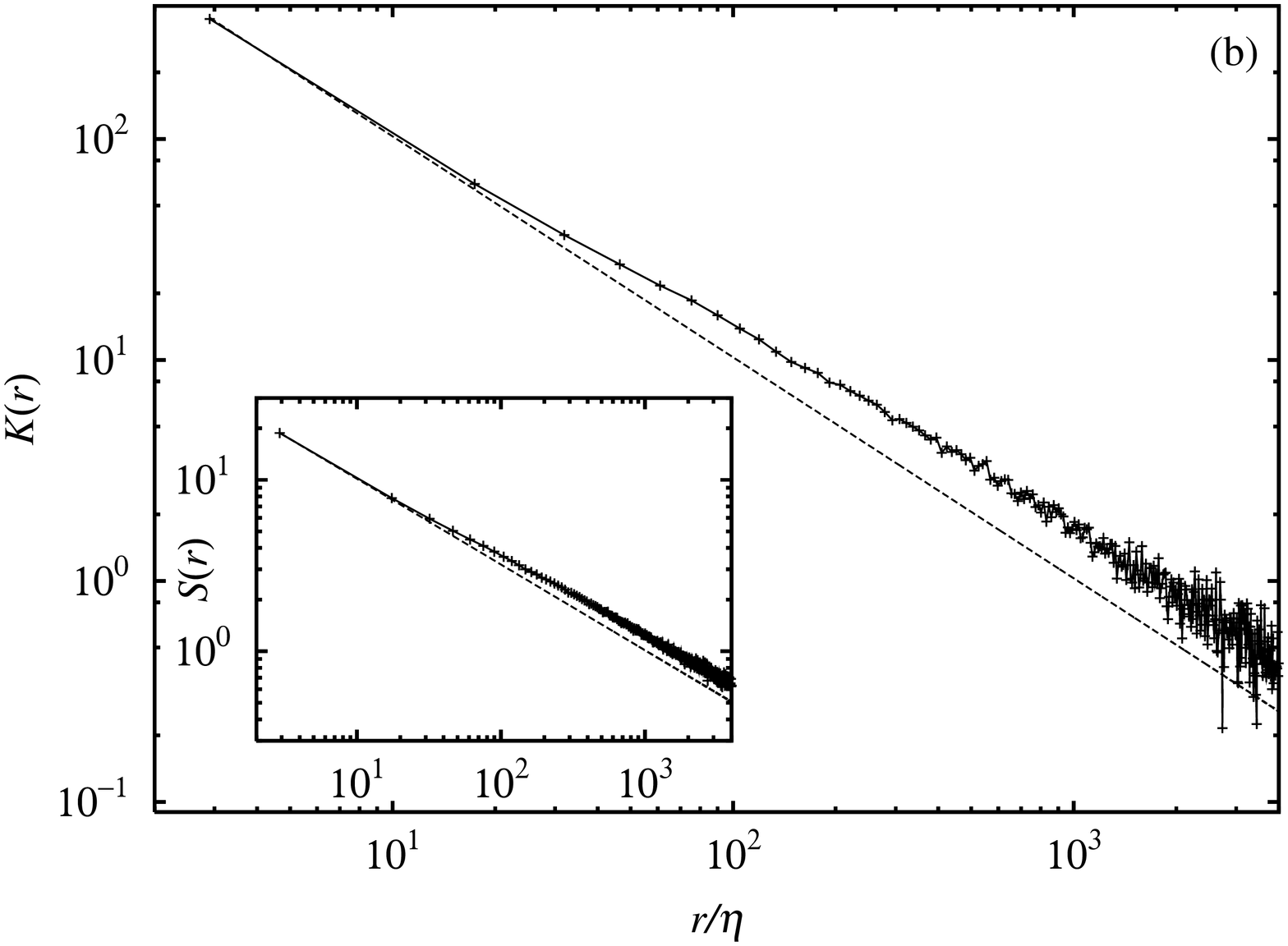,width=.3\textwidth,angle=-90}
\vspace*{0.3cm}
\caption{
Scale dependent Kurtosis, $K(r)$, for case
  (a) (top) and (b) (bottom).  Dotted lines represent the Poissonian
  behavior, that is $K_{(P)}(r) = \left( \rho\ r \right)^{-1}$. Notice
  that the Poisson scaling behavior is reached for large $r$-windows
  only scaling wise. In the insets the scale dependent Skewness,
  $S(r)$, behavior is shown. Again the Poissonian relation is drawn
  $S_{(P)}(r) = \left( \rho\ r \right)^{-1/2}$ (dotted line). \label{fig2}}
\end{center}
\end{figure}
In figure \ref{fig2} we show the Kurtosis and Skewness behavior,
evaluated for the two cases (a)-(b), in a comparison with the
Poissonian dependence. We observe in both cases a clear departure at
small scale from the scaling implied by the global homogeneity, 
which is 
only
recovered at the large scale ($\gtrsim L_0 \simeq 500 \eta$) where
the data points falls roughly parallel to the Poisson line.  The
departure from the Poisson line, that is noticeable already at the
scales immediately above $ \Delta r_{min}$, is an indication that
bubbles form clusters even at the smallest scale we are able to
detect, that is even below $5 \eta$ for case (a) or $3 \eta$ for case
(b). We observe that for the less turbulent case, $(b)$, the departure
from the homogeneous scaling is less marked.  A comparison with
synthetic Poisson samples of an equivalent number of bubbles, that we
have tried, shows that the available statistics is
sufficient to validate the deviations from the homogeneity discussed so
far. Scale dependent deviation from Poisson distribution is an
evidence of the fact that the dispersed microbubbles are trapped
within the dynamical vortical structures of turbulence.  Furthermore,
we observe that gravity plays a minor role in this dynamics. In fact, 
on average the bubbles are swept down by the mean flow and  
$g \tau_{b}/u_{\eta} \sim O(1)$ (see Tab. \ref{tab1}), which implies that even the smallest vortical structures of the flow may trap bubbles \cite{wang93}. Therefore, it is mainly the inertia that drives the bubble accumulation in the flow.

\section{Results of the analysis on numerical data}
To give further evidence for  the robustness
of the suggested
 statistical analysis of the hot-wire time series,  
we now
repeat the very same procedure with
 numerical simulation data.
We employ standard numerical tools already described and discussed
in details in \cite{maz03a,maz03b}.
In short,
we integrate
Lagrangian point-wise bubbles evolving on the background of an Eulerian
turbulent field.
The equation for the evolution of the point-wise bubble is the
following: \be \frac{d \bv}{dt} = 3 \frac{D \bu}{dt} -
\frac{1}{\tau_b} \lp \bv - \bu \rp - 2 \bg - \lp \bv - \bu \rp \times
\bomega 
\label{eq1} 
\ee where $\bu$ and $\bomega$ are respectively the fluid
velocity and vorticity computed at the bubble position and constitute
a simplified version of the model suggested in \cite{maxeyriley83}. 
The Eulerian flows is a turbulent homogenous and isotropic field
integrated in a periodic box, of resolution $128^3$, seeded
approximately with $10^5$ bubbles, corresponding to a void fraction
$\alpha = 4.5 \%$. Since previous experimental and numerical studies
\cite{ramon06b,maz03a} have revealed that the effect of bubbles on
strong unbounded turbulence is relatively weak, our numerical bubbles
are only coupled in one-way mode to the fluid, i.e. bubbles do not
affect the fluid phase.  The bubble-Reynolds number $Re_b$ is set to
unity and the Stokes number is $St \ll 1$. Therefore, the bubble radius is of order $\eta$, 
and the bubble terminal velocity $v_T = 2 g \tau_b$ in still fluid is smaller than
the smallest velocity scale $u_{\eta}$.
As (and actually even 
more than) in the experiment, the role of gravity is marginal.
\begin{table*}
\begin{center}
\begin{tabular}{|c|c|c|c|c|c|c|c||c|c|c|c|c|}
  \hline
  &$L_0$& $u'$  &$Re_{\lambda}$ &$\tau_{eddy}$  & $\tau_{\eta}$
  & $\eta$ & $u_{\eta}$ &$Re_b$& $R_0/\eta$ & $St$ & $g \tau_b/u_{\eta}$\\
  \hline
  (a')&
  5.0 & 
  1.4 &
   94 & 
   3.6 & 
  0.093 &
  0.025 &
  0.275 &
 1.0 &
 1.13 & 
  0.14 &
  0.55 \\
  \hline
  (b') &
   5.0 &
  1.0 &
  87 &
  4.9 &
  0.147 &
  0.032 &
  0.218 &
  1.0 &
  0.89 &
  0.09 &
  0.69\\
  \hline
  \hline
  &$L_0$ $(cm)$& $u'$ $(cm/s)$ &$Re_{\lambda}$ &$\tau_{eddy}$ $(ms)$ & $\tau_{\eta}$ $(ms)$
  & $\eta$ $(\mu m)$& $u_{\eta}$ $(cm/s)$&$Re_b$& $R_0/\eta$ & $St$ & $g \tau_b/u_{\eta}$\\
   \hline
  (a')& 0.41 & 7.2 & 94 & 57.3 &  4.7 & 68.7 & 1.45 & 1.0 & 1.13 & 0.14 & 0.55 \\
  \hline
  (b')& 0.46 & 5.5  & 87 & 82.5 & 7.3 & 85.7 & 1.16 & 1.0 & 0.89 & 0.09 &  0.69 \\
  \hline
\end{tabular}
\end{center}
\caption{Relevant turbulent scales and bubble characteristics for the two numerical   simulation performed. The top part reports the actual values in numerical  units from the simulation, the bottom part shows for comparison the corresponding physical equivalent quantities for air bubbles in water, this is to better appreciate similarities/differences with the experimental conditions of Table \ref{tab1}. The values on the bottom part are computed 
starting from the dimensionless quantities $Re_{\lambda}$, $Re_b$, $St$, and by assuming $\nu = 10^{-6}$ $m^2\ s^{-1}$ and $g = 9.8$ $m\ s^{-2}$.}
\label{tab2}
\end{table*}
In Table \ref{tab2} we report details of the numerical simulations,
these are chosen trying to match the experimental numbers. However, we
could not reach the same scale separations as in the experiments. In
the bottom panel of Tab.\ref{tab2} we translated the numerical units
to their physical equivalent. We note that in the numerics the Stokes
number, $St=\tau_b/\tau_{\eta}$, which is an indicator of the degree
of bubble interaction with turbulence, can not be as low as in the
experiments.
To achieve the same $St$ would require too much CPU time.
For practical reason the Stokes values adopted in our numerics are
roughly one order of magnitude larger than in the experiments,
although always much smaller than unity, $St \ll 1$.  Under this
conditions,  simple spatial visualization \cite{ber06b}
 show strong bubble
accumulation in nearly one-dimensional elongated structures in
correspondence to high enstrophy regions (identified as vortex
filaments.
\begin{figure}
\begin{center}
\vspace*{0.4cm}
\epsfig{file=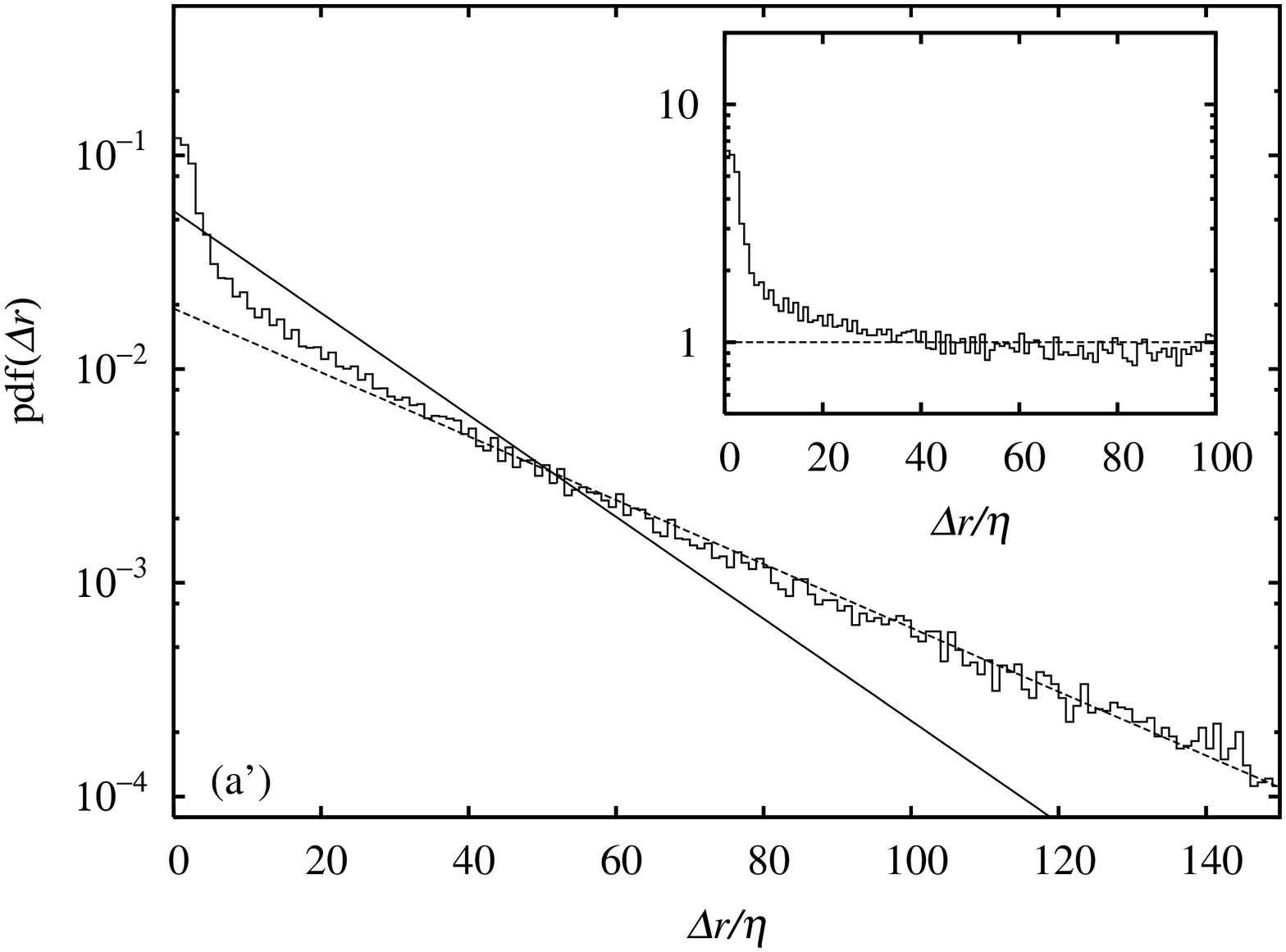,width=.3\textwidth,angle=-90}\\
\vspace*{0.6cm}
\epsfig{file=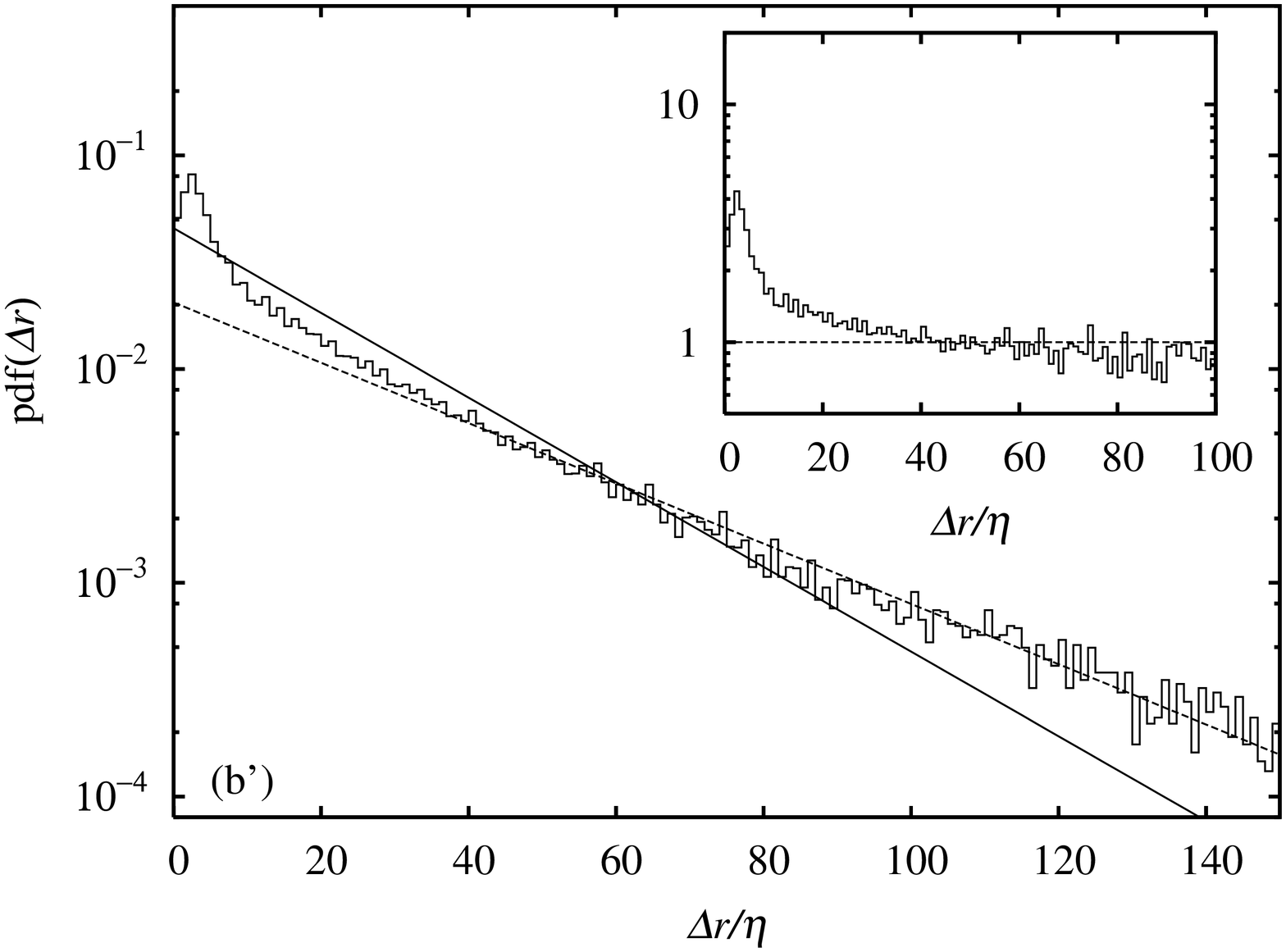,width=.3\textwidth,angle=-90}
\vspace*{0.3cm}
\caption{
Numerical result on the probability density
  function of distance between successive bubbles, pdf($\Delta
  r$). Case (a') (top) is the most turbulent. In the inset the same
  compensated plot as in Fig. \ref{fig1}}\label{fig3}
\end{center}
\end{figure}
As already stated, our goal is to use the numerics to confirm the
behavior of the suggested observables. To this end, we put $128$
virtual pointwise probes in the flow and recorded the hitting times of
the
bubbles, which we give a virtual radius $R_0$. 
The bubble radius is related to the bubble response time
$\tau_b$, namely $\tau_b = R_0 \equiv \lp 9 \tau_b \nu \rp^{1/2}$
when assuming no-slip boundary conditions at the gas-liquid interface.

An important difference between the experiments and the numerics is
the mean flow: It is present 
 in the experiment while intrinsically suppressed
in the simulations. In the numerical simulations the time is connected
to space displacements through the relation $\Delta R = \Delta
t\cdot u'$, where $u'$ is the root mean square velocity.
\begin{figure}
\begin{center}
\vspace*{0.4cm}
\epsfig{file=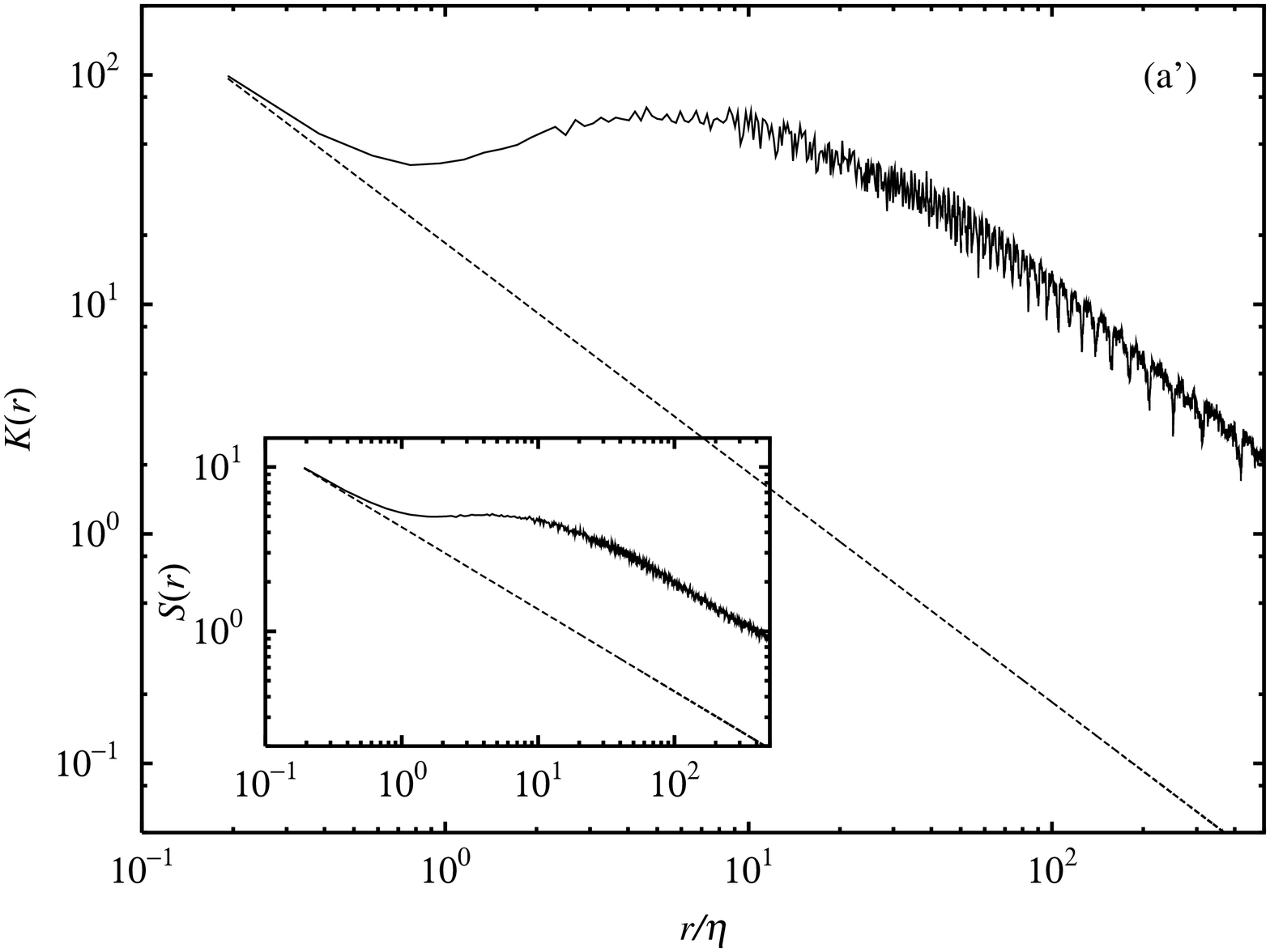,width=.3\textwidth,angle=-90}\\
\vspace*{0.9cm}
\epsfig{file=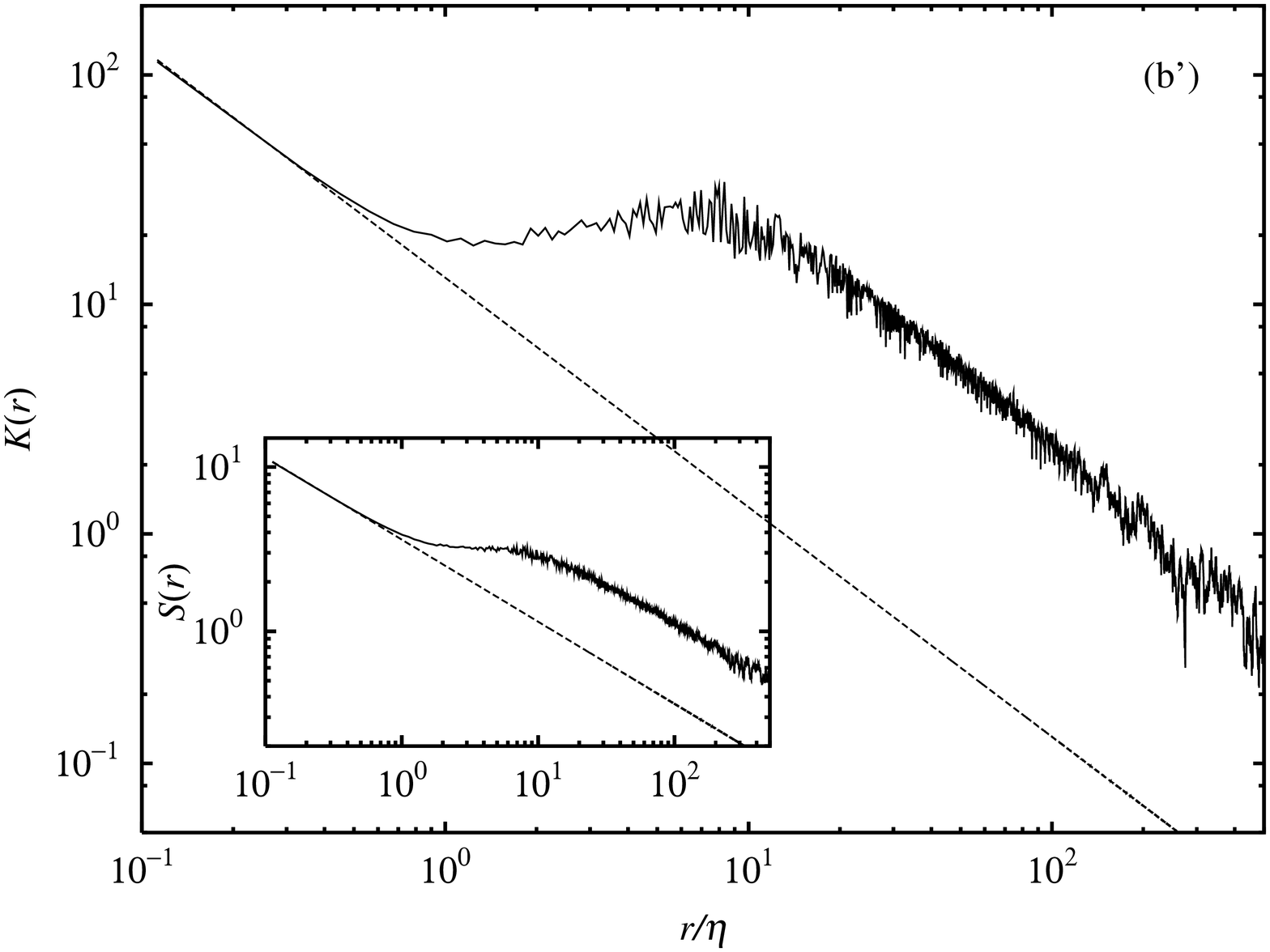,width=.3\textwidth,angle=-90}
\vspace*{0.6cm}
\caption{
Numerical result on scale dependent kurtosis,
  $K(r)$, for case (a') (top) and (b') (bottom), and Poissonian
  behavior (dotted). In the insets the scale dependent skewness,
  $S(r)$, behavior is shown.}\label{fig4}
\end{center}
\end{figure}
The level of turbulence, given the available resolution, has been
pushed as high as possible ($Re_{\lambda} \simeq 90$) to obtain a
better analogy with the experiment. Also in the numerical simulations
two cases with different Reynolds numbers are considered, see again Table \ref{tab2}.\\ \indent

In Figures \ref{fig3} and \ref{fig4} we show the results of the
statistical analysis of clustering from the time series obtained from
the numerical virtual probes. These two figures should be compared
with the analogous experimental findings already discussed and shown
in Figures \ref{fig1} and \ref{fig2}. Some qualitative similarities
are striking. First, starting from Fig. \ref{fig3}, we observe that
deviations from random and homogeneous, i.e. pure exponential
behavior, are relevant at small scales. This feature is confirmed by
the scale dependent kurtosis and skewness of Fig. \ref{fig4}, where
departure from the Poisson scaling already starts below $\eta$
scale. Second, the most turbulent case is the most clusterized, (a')
($Re_{\lambda} \simeq 94$) more than (b') ($Re_{\lambda} \simeq 87$).
The evaluation of the fraction of clustered bubbles, based on the fit
of the pdf($\Delta r$) as in the experiment, gives the value $29\%$
for (a') and $37\%$ for (b'). Though the qualitative
behavior of the statistical indicators is the same, 
also some important
differences arise in this comparison. 
First of all, 
full homogeneity in the numerics seems to be recovered already
at scales
of order ${\cal O}(10 \eta)$, 
whereas in the experiments
if was only recovered at 
${\cal O}(100 \eta)$. Furthermore, the deviations
from Poisson distribution and the fraction of clustered bubbles are
definitely stronger in the numerics. There are several possible
interpretation for this mismatch, including
the possible incompleteness of the
employed  model eq.\ (\ref{eq1}):  
First, some physical effects have been
neglected: the fluid-bubble and the bubble-bubble couplings and
the associated finite size effects (in the present conditions bubbles
can overlap!). A second reason can be the different degree of bubble
interaction with turbulence, a quantity that is parametrized by the
Stokes number $St=\tau_b/\tau_{\eta}$. The estimated $St$ in the
experiment is roughly one order of magnitude smaller than in the
simulation. This corresponds to bubbles that react faster to the fluid
velocity changes and hence to bubbles that closely follow 
the fluid
particles and accumulate less. Such a trend is also confirmed by our
numerics. 
\section{Conclusions}
We have performed statistical tests in order to detect and
quantitatively characterize the phenomenon of preferential
bubble 
concentration from single-point hot-wire anemometer
measurements in the bulk of a turbulent channel.  Our tools clearly
show that the experimental records display bubble clustering.  The
fraction of bubbles trapped in such structures is indeed considerable
 and
can be estimated to be of the order of $10\%$.  The scale-dependent
deviations from random homogeneous distribution, that we associate to
typical cluster dimension, extends from the smallest detectable scale,
${\cal O}(\eta $), to scales in the low inertial range, ${\cal
  O}(100 \eta $). Accumulation of bubbles is enhanced by increasing
the turbulence intensity.  Comparison with present Eulerian-Lagrangian
simulations, where point-like bubbles strongly accumulate in vortex
core regions, shows similar qualitative features and trends.\\  \indent 
We hope that our explorative investigation
 will stimulate new dedicated experiments and numerical simulations
to further quantify the clustering dynamics as function of Reynolds number and 
particle size, type, and concentration. The challenge is to further
develop
and employ quantitative statistical tools to allow for a meaningful
comparison between experiment and simulations, in order to validate
the modeling of particles and bubbles in turbulent flow.\\ \indent 
\textit{Acknowledgment: } 
We thank Stefan Luther and GertWim Bruggert for extensive help with the experimental setup and the measurements, and Kazuyasu Sugiyama for useful discussions. 
The work is part of the research program of the Stichting
voor Fundamenteel Onderzoek der Materie FOM, which is financially
supported by the Nederlandse Organisatie voor Wetenschappelijk
Onderzoek NWO.  

\end{document}